\documentstyle[aps,twocolumn]{revtex}
\begin{document}
\title{Formation energy, stress, and relaxations of
low-index Rh surfaces}
\author{Alessio Filippetti and Vincenzo Fiorentini}
\address{INFM -- Dipartimento di Scienze Fisiche, Universit\`a di
Cagliari, I-09124 Cagliari, Italy}
\author{Kurt Stokbro,\cite{p_add} Riccardo Valente,\cite{padd}
 and Stefano Baroni}
\address{INFM --
 Scuola Internazionale Superiore di Studi Avanzati, I-34014 Trieste, Italy}

\maketitle
\begin{abstract}
 {\it Ab initio} local-density-functional-theory
calculations of formation energies,
surface stress, and multilayer relaxations are reported for
the  (111), (100), and (110) surfaces of Rh.
The study is performed using ultrasoft pseudopotentials and
plane waves  in a parallel implementation.
\end{abstract}

\section{INTRODUCTION}\label{sec:intro}

In this paper we report ab initio calculations of surface energies,
stress and relaxations of low-index Rh surfaces.
Our results
 provide,  among others,
 original information on surface stress and surface stress
anisotropy, confirming the
recent findings by Feibelman for Pd and Pt \cite{feib}.
This is a positive  test of the feasibility and accuracy of
such calculations within the pseudopotential method.

A proper understanding of the basic properties of clean metal
surfaces is an essential step forward in the study of
adatom diffusion and of catalysis
of basic chemical reactions  on metal surfaces,
both topics of obvious fundamental  and
 technological interest \cite{general}. Ab initio calculations on
transition metal surfaces can provide adsorption geometries,
 dissociation barriers,  and diffusion paths
for use within empirical methods (e.g.  kinetic Montecarlo),
 and deeper insight into the basic governing mechanisms, and are
therefore of the utmost importance.

With a view at studying diffusion barriers \cite{altri}
and molecular dissociation \cite{st95-0} on low-index surfaces of
transition metals, we have undertaken a series of ab initio calculations
on the surfaces of Rh. Besides its interest as a catalyst, good
experimental data exist for seldiffusion on Rh, in particular for its
(100) face \cite{kellog}. Here we present  surface formation energies,
surface stress, and multilayer relaxations of the clean Rh low-index
surfaces; work in progress on surface vacancy formation, homoadsorption
and self-diffusion on Rh surfaces will be presented elsewhere. Our
calculations were performed within local-density-functional-theory
\cite{dft}, using  ultrasoft
pseudopotentials \cite{vand} to describe ion-electron interaction,
a plane-waves basis,
and iterative diagonalization algorithms in an
efficient parallel implementation \cite{valente}.
A newly developed variant of the Vanderbilt scheme \cite{vand} has been
used, whereby norm conservation is
only released for those angular-momentum channels which would
otherwise generate very hard potentials (the $d$-channel  in the
present case)  \cite{st95-1}.

For bulk Rh,  the following computational parameters
were found to give  converged total energy differences
and structural properties: plane-wave cutoff of 30 Ryd,
10 special-point  mesh for the fcc lattice, Fermi-surface smearing of
0.05 Ryd, first-order approximation of Methfessel and Paxton
\cite{mp} for the occupation numbers
distribution function (an up-to-date, lucid discussion of this
treatment of the metallic state in ab initio calculations has been
given recently by de Gironcoli \cite{degi}). For all
calculations, we used downfolded meshes equivalent to
the one used for the bulk.
The resulting bulk parameters are
$a_0=7.215$ bohr and $B = 3.079$ Mbar.
The theoretical  lattice constant was used in all calculations.
The surface areas per atom are
26.03 bohr$^2$, 22.54 bohr$^2$ and
36.81 bohr$^2$ for the (100), (111) and (110) surfaces, respectively.
Our results, given below in  eV/atom,  should be divided
by the  above values to switch to eV/bohr$^2$.

\section{RELAXED SURFACE GEOMETRIES}\label{sec:geom}

The  topmost layers
 of most  clean  transition-metal surfaces are known
to relax inward \cite{stu0}.
The intralayer spacing between the first two top layers,
in particular, is appreciably reduced with respect to its bulk
value. Such a relaxation is accompanied by smaller relaxations of
the second and third layers, that  may be directed both inward or outward.

The magnitude of the relaxations depends markedly on surface  orientation.
It is common knowledge that  larger inward relaxations occur
for rougher surfaces (i.e. with lower atomic density).
Our results, reported  for the three low-index clean surfaces
of Rh in Table \ref{t1}, confirm this tendency:
the (110) surface shows the largest relaxations, the closest-packed
(111) the smallest.
Also, even the second and third layer in (110) relax noticeably,
while for (100) and (111) only the top-layers show an appreciable shift.

The results in Table \ref{t1}
 were   obtained with 7-layers slabs, whereby all layers
have been relaxed to their equilibrium position. The relaxation
pattern is insensitive to a further increase of the number of layer for the
(100) and (111) surfaces. For the (110), changes are non-zero
but marginal: using  a 9-layers slab,
$\Delta d_{12}$ is almost unchanged (--10.7 \%), and  $\Delta d_{23}$ and
 $\Delta d_{34}$ are only slightly reduced (3.5 \% and --1.0
\% respectively).
We emphasize that results in Table \ref{t1}  are
 variations in interlayer spacing, which include relaxation
contributions  from the layers above and below the one considered.
For the top layer these values
are equal to the deviation from the ideal position  {\it only}
when the underlying layers are kept fixed.

In Table \ref{t1}, our results are also compared with those of Methfessel
{\it et al.} \cite{met}, who used  the FP-LMTO method to calculate the
 top layer relaxations while keeping all other layers fixed.
While the relaxation trend as function of roughness is the same in both
calculations, a sizable  quantitative difference appears for the (110)
surface. To clarify this, we repeated our calculations
relaxing only the first surface layer, obtaining  a relaxation
of  --7.5 \%. The deviation from our full results
 is thus due to the neglect of
multilayer relaxation in Ref. \cite{met}.
Indeed, this relaxation mode is expected to play a role
on the (110) face, where first-layer and third-layer atoms
are nearest-neighbors. Our results show that the assumption
(suggested in Ref. \cite{fu} for bcc (100) surfaces) that
the top interlayer spacing remains unchanged irrespective of
the movements of the layers below, is inappropriate for the
fcc (110) surface.

We now briefly recall Pettifor's \cite{pettifor} model
for the trend systematization of
the basic  properties of transition-metal surfaces.
In transition metals, according to this  model,
the bulk equilibrium lattice constant results
from the competition of a negative (attractive) pressure due to bonding,
localized, and directional
$d$ states,  and of the repulsive pressure of the delocalized $sp$ ones
(which tend to decrease their kinetic energy). At the surface, $sp$
electrons can
spill out into vacuum, and the top layer is further driven inward
by the $d$ electron attraction. Within this model, one expects
the top-layer relaxation to follow a roughly parabolic trend as
a function of $d$ occupation. (This feature  was basically
confirmed by the DFT-LDA calculations of Methfessel {\it et al.}
\cite{met}, where further detailed discussions can be found.)
At the end of a transition series, the $s$-$d$ balance is reversed,
with the $s$ electrons providing cohesion and the $d$ states
functioning more and more as a closed shell as one approaches the
noble metals. The fcc structure is typical of this section of the
transition series. Rh is  peculiar in that it is the first
metal in the 4$d$ transition series found to be stable in the fcc
structure, while still exhibiting typical transition metal-like features
such as sizable inward relaxations, high Fermi level density of states
(mostly $d$ in nature), and the ensuing efficient and short-ranged
screening.

\section{SURFACE FORMATION ENERGIES AND WORK FUNCTIONS}\label{sec:sigma}

Surface roughness plays a key  role in determining
the general trends of formation energy, work function, and relaxations.
One could pick any of these  quantities  as a quantitative measure of
the roughness, since each of them exhibits well defined trends as a
function of roughness. Within the Smoluchowsky model \cite{smol},
a rough surface presents a smaller inward-oriented
electric dipole moment compared to a smoother surface
of the same material. One thus  expects the (110) surface to
have the largest
formation energy (having the highest number of removed nearest neighbors)
and the smallest work function; on the other hand, the close-packed (111)
surface should have the smallest surface energy and the
largest work function. Our results
for the surface energies $\sigma$
and work functions $W$ of the three surfaces, listed in Table \ref{t2},
are in line with these expectations.

Results are given for both relaxed and unrelaxed surfaces. Surface
energies and work functions are only marginally
affected by relaxations, even for the rough (110) surface.
The discussion  of this feature given in a previous work on W
\cite{fu} applies to Rh as well.
Our results are seen to compare well with those
of Ref. \cite{met}. Numerical deviations are quite minor,
considering the difference between the two methods employed.

 A technical  point concerning the calculation of surface energies
 is in order here.
 It is generally assumed that surface energies can be extracted
 straightforwardly as differences of the energy of a slab mimicking the
 surface, and of an appropriate bulk energy. For the present case
\begin{equation}
\sigma = \lim_{N\rightarrow\infty} \frac{1}{2} (E^{N}_{\rm slab} - N\,
 E_{\rm bulk}), \label{uno}
\end{equation}
with $N$ the number of layers and  $E_{\rm bulk}$ the bulk energy
per atom. It has been recently proven by Boettger \cite{boet}
 that the  surface energy  diverges
 as a function of slab thickness if the incremental energy of
 the slab upon addition of an atom
differs (however little) from the bulk energy
$E_{\rm bulk}$.
While this results has not yet been  widely appreciated,
in a recent work it has been shown \cite{fm} that the problem
is  indeed relevant in practice even for technically state-of-the-art
calculations. A simple solution to the problem was suggested
 \cite{fm} : the bulk energy to be used in Eq. \ref{uno} is
the linear part of the slab total energy as a function of
$N$. This choice of $E_{\rm bulk}$, besides being the
natural one, leads to fast convergence of
$\sigma$ (see Ref. \cite{fm} for further details).
In our calculations we indeed encountered essentially the same
situation mentioned in Ref. \cite{fm} for Pt (100),
and  computed $\sigma$ by the method suggested there.

\section{SURFACE STRESS}\label{sec:stress}

Surface stress is an
  important quantity providing insight into
surface structure, and useful trend information on
a number of processes, such as surface
recontructions.  For example, relief of tensile in-plane stress
 is believed to  cause the quasi-hex
reconstruction of  the (100) surface of $5d$ fcc transition metals \cite{fms};
 a contractive reconstruction observed  recently on Cu (100)
\cite{muller} has been attributed to the same cause \cite{me}.
The  surface stress is defined as the
strain derivative of surface energy per surface cell \cite{feib,vand2}.
If all  bulk contributions to the stress are zero, the surface stress is
thus
\begin{equation}
\tau_{\alpha\beta}^{\rm surf} \equiv
 {1 \over A}\>{\partial \sigma \over \partial
\epsilon_{\alpha\beta}}
= \frac{\Omega}{2 A} \tau_{\alpha\beta},
\end{equation}
where  $\tau_{\alpha\beta}$ is the volume-averaged
stress tensor \cite{niels} of the supercell.
As usual the factor of 2 accounts for the two surfaces of the slab.
A positive
$\tau^{\rm surf}_{\alpha\beta}$
is  a tensile stress favoring  in-plane contraction
of the surface, while a negative (compressive) stress favors surface
expansion.

For the stress calculations, we used an energy cutoff of
45 Ryd, at  which the bulk stress is
essentially zero ($\tau_{\alpha\beta}^{\rm bulk} \simeq
5 \times 10^{-6}$ eV/bohr$^3$), and dispensed
therewith with bulk corrections to the surface stress \cite{vand2}.
If the bulk stress were  non zero,  it should  be subtracted out much
in the same  way as in the calculation of the surface energy.
In passing we verified that  the surface energies calculated at 45 Ry
deviate from those calculated at 30 Ry by less than 0.01 eV/atom.

In Table \ref{t3} we report the surface stress for the
(111), (100) and (110) surfaces of Rh.
For the unrelaxed surface the planar components of the stress
are sizably larger than the vertical component. The latter is
 however non-zero, and in fact quite large for the very open (110) surface.
Relaxation reduces appreciably the in-plane components (up to about 50 \%)
and renders  the $z$ component of the supercell stress negligible in
comparison to the planar components, as it should be.

As expected, the stress is anisotropic for the (110) surface. The
ratio of the stress components along the [1$\overline{1}$0] and
[001] directions is  1.59 (1.09) for the relaxed (unrelaxed) surface.
These appear consistent with the anisotropy values of
 1.47 (1.05) for relaxed (unrelaxed)  Pd (110),
and 2.08 (1.47) for relaxed (unrelaxed) Pt (110)
recently reported by Feibelman \cite{feib}.
Both  anisotropy and absolute values
of the (110) stress components are comparable to
those of Pd, but much smaller than for Pt. This seems
consistent with the absence of missing-row reconstructions of clean
Rh (110) and Pd (110), and its  presence for Pt (110).
Note also that anisotropy is strongly enhanced by relaxation due to
a larger decrease of the stress along [001] (i.e.
transversally to the surface channels),
again in agreement with Feibelman's results
for Pt and Pd.
Also, the stress values are quite insensitive to the number of
layers used in its calculation, which is consistent with Feibelman's
finding that close to 99 \% of the stress on a Pt (111) surface
originates from the top surface layer.

To our knowledge these are the first calculations of stress for
Rh surfaces. The only  other stress calculation for Rh we are aware
of is that of Ref. \cite{fms}, in which a stress of 1.94 eV/atom was
obtained for unrelaxed Rh (100). This compares well with the value of
2.03 eV/atom obtained here, expecially in view of the very different
computational methods  used in the two cases.

\section{SUMMARY}

The present ab initio calculations of  surface formation
energies, work functions, relaxations,
and stress for the clean low-index surfaces of Rh provide an
encouraging test of the accuracy of the (ultrasoft)-pseudopotential
 method for transition metal surface studies.
 It is anticipated that it  will soon be  possible
to study e.g. the surface stress of reconstructed surfaces,
which should clarify the role of stress and stress anisotropy
in surface reconstructions. The study of adsorption and
 of simple surface defects (vacancies, adatom-vacancy pairs),
and diffusion processes, also seems within reach.

\section{Acknowledgements}

We thank Sabrina Oppo for useful discussions and help with the
k-point folding code.
The Computing Center of CRS4 (Centro Ricerche e Studi
Superiori in Sardegna), Cagliari,
provided access to, and computing time on an IBM SP2
within a collaborative framework between CRS4 and the University
of Cagliari.


\begin{table}[ht]
\centering\begin{tabular}{l|rrr|rrr}
  &\multicolumn{3}{c}{This work} &\multicolumn{3}{c}{Ref. \cite{met}}  \\
\tableline
  & $(100)$ & $(110)$ & $(111)$& $(100)$ & $(110)$ & $(111)$ \\
\tableline
$\Delta d_{12}$  &--3.4 &--10.5 &--1.6 &--3.5 &--7.5 &--2.5  \\
\tableline
$\Delta d_{23}$ & 0.5 &4.4 & --0.4& & & \\
\tableline
$\Delta d_{34}$ &--0.3 &--1.6 & 0.3& & & \\
\end{tabular}

\caption[T1]{\footnotesize Intra-layer relaxations (percentage
variation with respect to ideal layer
spacing) for the three low-index surfaces of Rh.
Results from Ref. \cite{met} are included for comparison.}
\label{t1}
\end{table}

\begin{table}[ht]
\centering\begin{tabular}{l|rrr|rrr}
  &\multicolumn{3}{c}{This work}
&\multicolumn{3}{c}{Ref. \cite{met}} \\
\tableline
  & $(100)$ & $(110)$ & $(111)$& $(100)$ & $(110)$ & $(111)$ \\
\tableline
$\sigma^u$  &1.34  &1.96 &0.98 &     &     &     \\
\tableline
$\sigma^r$  &1.32  &1.89 &0.97 &1.27 &1.84 &0.99 \\
\tableline
$\Delta \sigma$  &0.02  &0.07 &0.01 &     &     &     \\
\tableline
$W^u$      &5.45  &5.07 & 5.59 &     &     &     \\
\tableline
$W^r$      &5.46  &5.07 & 5.56 &5.25 &4.94 &5.44 \\
\end{tabular}

\caption[T2]{\footnotesize Surface energy
$\sigma$, relaxation energy $\Delta \sigma$ (the surface energy
change upon relaxation) and work function
for the three clean surfaces.
'u' and 'r' indicate unrelaxed and
relaxed surfaces respectively. All results are in eV/atom.
Results of Ref. \cite{met} are included for comparison.}
\label{t2}
\end{table}

\begin{table}[ht]
\centering\begin{tabular}{l|rr}
  & $\tau_{xx}^{\rm surf}$
& $\tau_{yy}^{\rm surf}$ \\
\tableline
(111) u   &\multicolumn{2} {c}{1.46}  \\
\tableline
(100) u   &\multicolumn{2} {c}{2.03}  \\
\tableline
(110) u   & 2.65 & 2.87  \\
\tableline\tableline
(111) r   &\multicolumn{2} {c}{1.17}   \\
\tableline
(100) r   &\multicolumn{2} {c}{1.43}   \\
\tableline
(110) r   & 1.25 & 2.01 \\
\end{tabular}

\caption[T3]{\footnotesize Surface stress (eV/atom)
for unrelaxed ('u') and relaxed ('r') Rh
surfaces. For the (110) surface, the $xx$   component is the
 [001] one, and the $yy$ is the  [1$\overline{1}$0].
See the text for the effects of relaxation.}
\label{t3}
\end{table}
\end{document}